\documentclass{iau} 
\usepackage{units}
\usepackage{natbib}
\usepackage{graphicx}

\title[Split main sequence in the open cluster NGC 2287] 
{Stellar rotation bifurcation caused by tidal locking in the open cluster NGC 2287?}

\author[W. Sun \textit{et al.}]   
{Weijia Sun$^{1,2}$, Chengyuan Li$^{3,2}$, Licai Deng$^{2,4,5}$ \and Richard de Grijs$^{6,7,8}$
}

\affiliation{$^1$Kavli Institute for Astronomy \& Astrophysics and Department of Astronomy, Peking University, Yi He Yuan Lu 5, Hai Dian District, Beijing 100871, China \\ email: {\tt swj1442291549@gmail.com} \\[\affilskip]
$^2$Key Laboratory for Optical Astronomy, National Astronomical Observatories, Chinese Academy of Sciences, 20A Datun Road, Chaoyang District, Beijing 100012, China \\[\affilskip]
$^3$School of Physics and Astronomy, Sun Yat-sen University, Zhuhai 519082, China \\[\affilskip]
$^4$School of Astronomy and Space Science, University of the Chinese Academy of Sciences, Huairou 101408, China \\[\affilskip]
$^5$Department of Astronomy, China West Normal University, Nanchong 637002, China \\[\affilskip]
$^6$Department of Physics and Astronomy, Macquarie University, Balaclava Road, Sydney, NSW 2109, Australia \\[\affilskip]
$^7$Research Centre for Astronomy, Astrophysics and Astrophotonics, Macquarie University, Balaclava Road, Sydney, NSW 2109, Australia \\[\affilskip]
$^8$International Space Science Institute--Beijing, 1 Nanertiao, Hai Dian District, Beijing 100190, China \\[\affilskip]}

\pubyear{2019}
\volume{351}  
\jname{Star Clusters: From the Milky Way to the Early Universe}
\editors{A. Bragaglia, M.B. Davies, A. Sills \& E. Vesperini, eds.}
\begin{document}

\maketitle

\begin{abstract}
We present a detailed analysis of the projected stellar rotational
velocities of the well-separated double main sequence (MS) in the
young, $\sim\unit[200]{Myr}$-old Milky Way open cluster NGC 2287 and
suggest that stellar rotation may drive the split MSs in NGC 2287. We
find that the observed distribution of projected stellar rotation
velocities could result from a dichotomous distribution of stellar
rotation rates. We discuss whether our observations may reflect the
effects of tidal locking affecting a fraction of the cluster's member
stars in stellar binary systems. The slow rotators are likely stars
that initially rotated rapidly but subsequently slowed down through
tidal locking induced by low-mass-ratio binary systems. However, the
cluster may have a much larger population of short-period binaries
than is usually seen in the literature, with relatively low secondary
masses.  
\end{abstract}

\firstsection 
\section{Introduction}

Discoveries of extended main-sequence (MS) turnoffs (eMSTOs) in
intermediate-age clusters \citep[e.g.][]{2008ApJ...681L..17M,
  2009A&A...497..755M} and split MSs in young clusters
\citep[e.g.,][]{2015MNRAS.450.3750M, 2015MNRAS.453.2637D} have
generated great interest in the formation mechanisms of these
phenomena in clusters. These features, which are primarily found in
the Magellanic Clouds, strongly challenge our understanding of star
cluster formation and evolution.

Multiple scenarios have been proposed to explain the observational
results, including an extended star formation history
\citep{2011ApJ...737....3G}, variable stars
\citep{2016ApJ...832L..14S}, stellar rotation, and convective
overshooting \citep{2017ApJ...836..102Y}. Thus far, the stellar
rotation interpretation is most favored by a range of observations,
including the detection of a population of Be stars based on
narrow-band photometry \citep{2017MNRAS.465.4795B,
  2018MNRAS.477.2640M} and rapidly rotating stars based on direct
spectroscopy \citep{2017ApJ...846L...1D, 2018AJ....156..116M},
although contributions from other channels cannot be completely ruled
out.

Recent studies have revealed that these features are not exclusive to
Magellanic Clouds clusters, but they are quite common in Galactic open
clusters \citep{2018ApJ...869..139C, 2019ApJ...876..113S}. These
results provide us with a great opportunity to study cluster evolution
under different physical conditions. The cluster we selected to study
is NGC 2287. It is an open cluster that is known to have an
eMSTO. This study aims to verify whether there is a connection between
the MS stars' loci in the color--magnitude diagram (CMD) and their
stellar rotation properties, and study the formation mechanism(s)
responsible for its peculiar rotation distribution.

\section{Stellar rotation and split MSs in NGC 2287}

\begin{figure}[b]
\begin{center}
\includegraphics[width=2.2in]{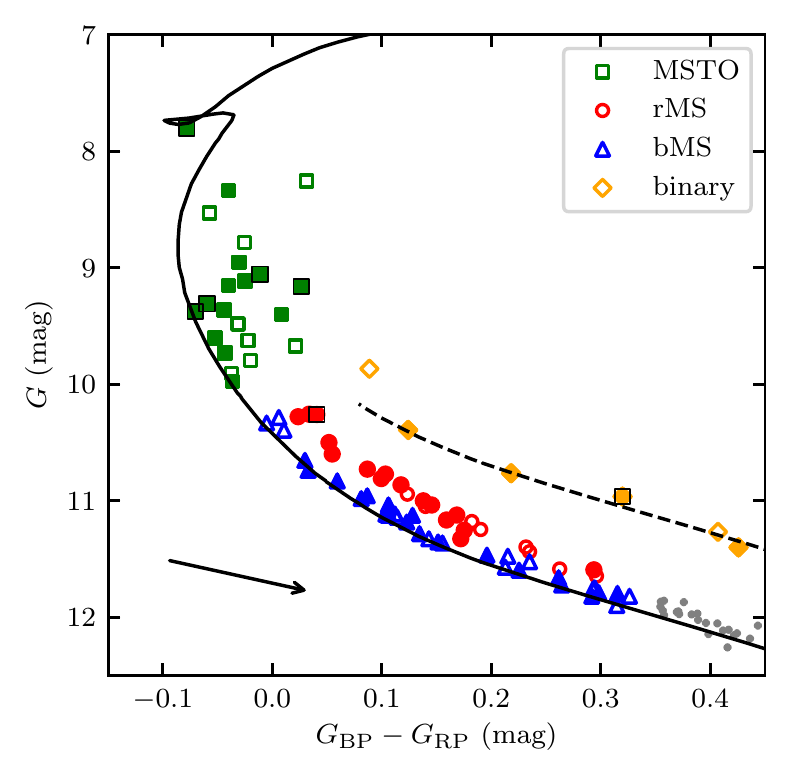}
\includegraphics[width=2.2in]{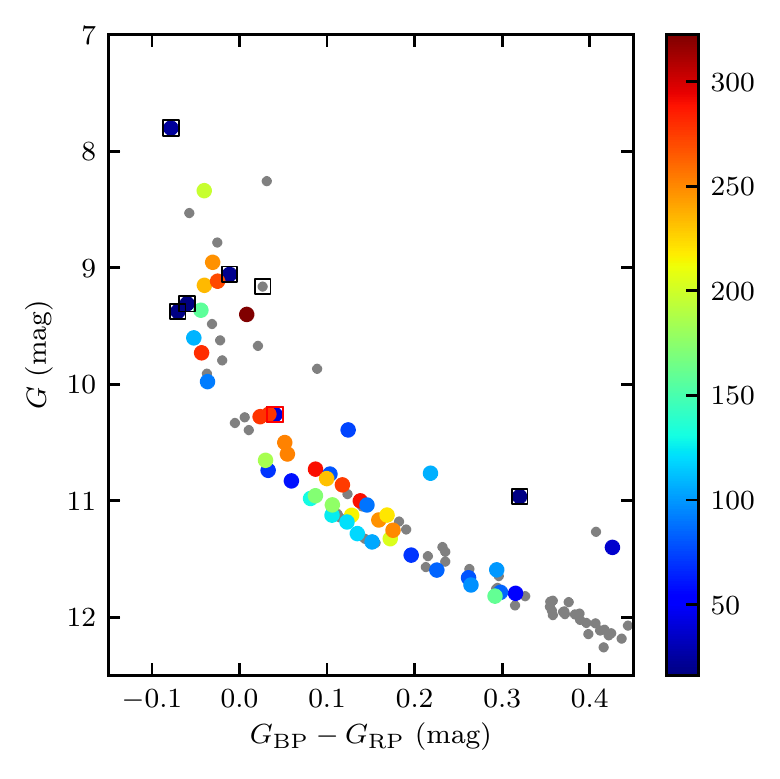}
\caption{(left) CMD of NGC 2287 around its eMSTO and split-MS
  regions. bMS, rMS, MSTO, and binary stars are represented by blue
  triangles, red circles, green squares, and yellow diamonds,
  respectively. Spectroscopically analyzed stars are marked with solid
  symbols. The black solid and dashed lines represent the best-fitting
  {\sc parsec} isochrone and the corresponding equal-mass binary
  sequence, respectively. The black arrow indicates the direction of
  the reddening vector. (right) CMD of NGC 2287 with stars color-coded
  according to their projected rotational velocities. Squares
  represent spectroscopic binaries. Adapted from Sun et al. (submitted to ApJ)\label{fig:cmd}}
\end{center}
\end{figure}

Following standard procedures to select member stars using proper
motion and parallax analysis, we derived a `clean' membership
determination for NGC 2287 based on \textit{Gaia} DR2. The
\textit{Gaia}-based CMD of the cluster's member stars is presented in
the left-hand panel of Fig.~\ref{fig:cmd}. A clearly defined split MS
is shown for $G$-band magnitudes from \unit[10.5]{mag} to
\unit[11.5]{mag}. We also overplotted the best-fitting isochrone with
an age of \unit[150]{Myr} based on the {\sc parsec 1.2s} models
\citep{2012MNRAS.427..127B} for solar metallicity, $Z = 0.0152$, and
an extinction $A_\mathrm{V} = \unit[0.217]{mag}$. By checking the MS
region that is not parallel to the reddening direction, we confirmed
that the split MS and eMSTO in NGC 2287 are not artifacts caused by
differential extinction.

We downloaded archival high-resolution ($R\approx 29,000$) spectra
observed with the European Southern Observatory's (ESO) Very Large Telescope (VLT)
equipped with the FLAMES/GIRAFFE spectrograph. The data were collected
as part of program 380.D-0161 (PI: Gieles). We studied 53 bright stars
of the full 166 member-star sample observed, covering the eMSTO, the
red and blue MSs (rMS, bMS), and the equal-mass binary sequence (see
Fig.~\ref{fig:cmd}, left). Using the absorption-line profiles of the
Mg{\sc i} triplet, we calculated the projected rotational velocities,
$v\sin i$, by comparing the observed spectra with synthetic stellar
spectra convolved with various rotational velocities from the Pollux
database \citep{2010A&A...516A..13P}. As for stars with multiple
observations, we confirmed that most do not show any variations in
their stellar parameters, including radial and rotational
velocities. There are a handful of exceptions that may be attributed
to spectroscopic binaries (black squares in Fig.~\ref{fig:cmd}).

The right-hand panel of Fig.~\ref{fig:cmd} presents the CMD of NGC
2287 with stars color-coded according to their $v\sin i$. bMS and rMS
stars are clearly separated in projected rotational velocity, in the
sense that bMS stars are mainly dominated by slow rotators while rMS
stars are rapid rotators. The mean projected rotational velocities of
bMS and rMS are $\langle v\sin i\rangle_\mathrm{bMS} =
\unit[111\pm13]{km\,s^{-1}}$ ($\sigma = \unit[46]{km\,s^{-1}}$) and
$\langle v\sin i\rangle_\mathrm{bMS} = \unit[255\pm10]{km\,s^{-1}}$
($\sigma = \unit[26]{km\,s^{-1}}$), respectively, if possible unresolved
binaries are excluded. A similar trend is observed for MSTO stars,
i.e., stars are getting redder as their rotation rates increase.

\section{Discussions}

To unravel the underlying distribution of equatorial rotational
velocities, $v_\mathrm{eq}$, and the inclination angles, $i$, we
compared the cumulative distribution of the projected rotational
velocities $v\sin i$ with three distribution models. Model 1 has a
uniform distribution of both $v_\mathrm{eq}$ and orientation in
three-dimensional space. Model 2 is characterized by a uniform
distribution of $v_\mathrm{eq}$ and a Gaussian distribution of $i$,
which is the representative scheme for `spin alignment'
\citep[see][]{2017NatAs...1E..64C, 2018A&A...618A.109M}. Model 3
adopts a uniform distribution of $i$ and a bimodal distribution of
$v_\mathrm{eq}$ to represent the slowly and rapidly rotating
populations, $v_\mathrm{s}$ and $v_\mathrm{r}$, respectively.

We found that model 3 performs significantly better than the other two
models. It yields two rotating populations with peak velocities of
$v_\mathrm{r} = \unit[280]{km\,s^{-1}}$ and $v_\mathrm{s} =
\unit[100]{km\,s^{-1}}$. These values are within $\unit[1]{\sigma}$ of
$\langle v\sin i\rangle_\mathrm{rMS}$ and $\langle v\sin
i\rangle_\mathrm{bMS}$, suggesting a relationship between fast/slow
rotators and rMS/bMS stars. This represents the first evidence in
support of a dichotomous distribution of real rotational velocities in
star clusters. It raises an important question as to the origin of
such a bimodal rotational velocity distribution.

Although a similar bimodal distribution of rotation has been detected
in early-type field stars \citep{2007A&A...463..671R,
  2012A&A...537A.120Z}, there are substantial differences in terms of
mass and velocity. On the one hand, the census of field stars found a
bimodal distribution only for stars more massive than
$\unit[2.5]{M_\odot}$, while the lowest-mass end of the split MS in
NGC 2287 is around $\unit[1.7]{M_\odot}$. On the other hand, the peak
rotational velocities we derived for slow and fast rotators in NGC
2287 are both slightly larger than the values reported by
\citet{2012A&A...537A.120Z} for a similar mass range.

\citet{2017NatAs...1E.186D} proposed a possible scenario that could
explain the bimodal rotational distribution in NGC 2287. They argued
that bMS in young clusters might be the outcome of fast braking of the
rapidly rotating population. Magnetic-wind braking or tidal torques
owing to a binary companion can, in theory, rapidly decelerate a
star's rotation rate and transfer the star's evolution from the
rapidly rotating to the non-rotating track. Here, we explore whether
the bMS stars in NGC 2287 could be composed of a population of slow
rotators ($\sim \unit[100]{km\,s^{-1}}$) which may have slowed down
from their initial state of rapid rotation by low-mass-ratio ($q \le
0.4$) binary companions.

Using the equations given by \citet{2002MNRAS.329..897H}, we estimated
the synchronization timescale for a typical star in the split MS
region of NGC 2287. We confirmed that the synchronization timescale of
a close binary system is relatively short compared with the age of NGC
2287. And the possible distribution of binary separations, $a$, and
mass ratios, $q$, overlap significantly with the expected relation
between $a$ and $q$ for slow rotators, based on Kepler's Third
Law. Such a population of close binaries with low-mass-ratio
companions could explain the existence of the bMS. In the meantime,
the higher-mass-ratio ($q\geq 0.5$) unresolved binaries will become
significantly reddened and therefore appear on the rMS rather than the
bMS. However, there are some possible flaws in the scenario. The
best-fitting result of Model 3 requires a bMS-to-rMS number ratio of
close to unity, which is not found in the field survey
\citep{2017ApJS..230...15M}. Another peculiar feature is the mass
ratio distribution of NGC 2287. The well-defined equal-mass binary
sequence implies that they have mass ratios close to unity. Our
observational evidence shows that in NGC 2287 binary systems have
either very low or very high (near unity) mass ratios since no objects
with intermediate-mass ratios appear to be present in the CMD.

\section*{Acknowledgement}
R. d. G. and L. D. acknowledge research support from the National
Natural Science Foundation of China through grants 11633005, 11473037,
and U1631102. C. L. and L. D. are grateful for support from the
National Key Research and Development Program of China through grant
2013CB834900 from the Chinese Ministry of Science and Technology. This
work has made use of data from the European Space Agency (ESA) mission
\textit{Gaia} (https://www.cosmos.esa.int/gaia), processed by the
\textit{Gaia} Data Processing and Analysis Consortium (DPAC,
https://www.cosmos.esa.int/web/gaia/dpac/consortium). Funding for the
DPAC has been provided by national institutions, in particular the
institutions participating in the \textit{Gaia} Multilateral
Agreement. It is also based on observations made with ESO telescopes
at the La Silla--Paranal Observatory under program ID 380.D-0161.

\end{document}